\documentclass[aps,prl,superscriptaddress,twocolumn,showpacs,preprintnumbers,amsmath,amssymb]{revtex4-1}

\usepackage{amsfonts}
\usepackage{amsthm}

\usepackage{graphicx}

\begin{document}

\title{Evidence for Triplet Superconductivity in a Superconductor-Ferromagnet Spin Valve}

\author{P.~V.\ Leksin}
\author{N.~N.\ Garif'yanov}
\affiliation{Zavoisky Physical-Technical Institute, Kazan Scientific Center of Russian Academy of Sciences, 420029 Kazan, Russia}

\author{I.~A.\ Garifullin}
\email{ilgiz\_garifullin@yahoo.com}
\affiliation{Zavoisky Physical-Technical Institute, Kazan Scientific Center of Russian Academy of Sciences, 420029 Kazan, Russia}

\author{Ya.~V.\ Fominov}
\affiliation{L.~D.\ Landau Institute for Theoretical Physics RAS, 119334 Moscow, Russia}
\affiliation{Moscow Institute of Physics and Technology, 141700 Dolgoprudny, Russia}

\author{J.\ Schumann}
\author{Y.\ Krupskaya}
\author{V.\ Kataev}
\author{O.~G.\ Schmidt}
\affiliation{Leibniz Institute for Solid State and Materials Research IFW Dresden, D-01171 Dresden, Germany}

\author{B.\ B\"{u}chner}
\affiliation{Leibniz Institute for Solid State and Materials Research IFW Dresden, D-01171 Dresden, Germany}
\affiliation{Institut f\"{u}r Festk\"{o}rperphysik, Technische Universit\"{a}t Dresden, D-01062 Dresden, Germany}

\date{12 September 2012}

\begin{abstract}
We have studied the dependence of the superconducting (SC)
transition temperature on the mutual orientation of magnetizations
of Fe1 and Fe2 layers in the spin valve system
CoO$_{x}$/Fe1/Cu/Fe2/Pb. We find that this dependence is
nonmonotonic when passing from the parallel to the antiparallel
case and reveals a distinct minimum near the orthogonal
configuration. The analysis of the data in the framework of the SC
triplet spin valve theory gives direct evidence for the long-range
triplet superconductivity arising due to noncollinearity of the
two magnetizations.

\centerline{\textbf{Published as: Phys.\ Rev.\ Lett.\ \textbf{109}, 057005 (2012)}}
\end{abstract}



\maketitle

Recent theories predict that  \emph{spin-triplet} components of the superconducting (SC) condensate can be generated under certain conditions in
heterostructures comprising a conventional superconductor (S) and a ferromagnet (F) (for a review see, e.g., \cite{bergeret3}). The components with
the spin projections (upon exchange field) $S_z=\pm 1$ can then penetrate into a ferromagnet over a long distance. The appearance of the triplet
components can be understood in the framework of the picture proposed by Demler \textit{et al.} \cite{Demler}: Upon entering the F layer, a pair of
electrons with spin up and spin down $\left| \uparrow \downarrow \right>$ acquires a center-of-mass momentum $Q=2h/v_F$, since the exchange field
acts on two electrons as a potential of opposite signs. Here, $h$ is the exchange field and $v_F$ is the Fermi velocity. Taking into account a
complementary pair with opposite spins, $\left| \downarrow \uparrow \right>$ (necessary for fermionic antisymmetry), that acquires the opposite
center-of-mass momentum $-Q$, one combines them into a singlet Cooper pair. The wave function of this Cooper pair becomes spatially modulated in the
F layer as $\cos (Qx)$, where $x$ is the center-of-mass coordinate. This is the reason for oscillating behavior of the S/F proximity effect
\cite{buzdin}. However, at that time, Demler \textit{et al.} did not pay attention to the triplet component which inevitably arises in the framework
of the same approach. Indeed, the singlet state of the two spins (corresponding to the singlet Cooper pair) is an antisymmetric combination
$\left|\uparrow\downarrow \right>-\left| \downarrow \uparrow \right>$ (hereafter, we omit the normalizing factor $1/\sqrt{2}$ for brevity). In the F
layer, these two terms acquire oscillating exponential prefactors differing by the sign of $Q$. This yields \cite{eschrig}
\begin{multline}
e^{iQx}\left|\uparrow\downarrow \right>-e^{-iQx}\left| \downarrow
\uparrow \right>\\
= \left( \left| \uparrow\downarrow \right>-\left| \downarrow
\uparrow \right> \right) \cos{Qx}+\left( \left| \uparrow\downarrow
\right>+\left| \downarrow \uparrow \right> \right)i \sin{Qx},
\end{multline}
where the cosine term is again the singlet component, whereas the sine term is the triplet component with zero $z$ projection (in the direction of
the exchange field). If we now introduce another \emph{noncollinear} direction of $h$, we can rotate the quantization axis accordingly, and then in
the new coordinate system we immediately obtain also the  components with $S_z=\pm 1$ (i.e., $\left| \uparrow\uparrow
\right>$ and $\left| \downarrow\downarrow \right>$). Those components are not destroyed by the exchange field and can penetrate deeply into the F
layer just as the singlet superconductivity penetrates into a nonmagnetic metallic layer; therefore, they are often referenced as ``long-range''
triplet components (LRTC).

In this Letter, we present experimental evidence for the occurrence of the LRTC of the SC condensate in the multilayer spin-valve heterostructure
CoO/Fe1/Cu/Fe2/Pb. It is manifested in the nonmonotonic behavior of the SC transition temperature $T_c$ of the Pb layer upon gradual rotation of the
magnetization of the ferromagnetic Fe2 layer $\boldsymbol M_\mathrm{Fe2}$ with respect to the magnetization of the Fe1 layer $\boldsymbol M_\mathrm{Fe1}$
from the parallel (P) to the antiparallel (AP) orientation. We observe a clear minimum of $T_c$ for the orthogonal configuration of $\boldsymbol
M_\mathrm{Fe1}$ and $\boldsymbol M_\mathrm{Fe2}$. As follows from our analysis in the framework of the theory of the SC triplet spin valve \cite{fominov},
such minimum of $T_c$ is a fingerprint of the spin-triplet component generated by noncollinear magnetizations $\boldsymbol M_\mathrm{Fe1}$ and
$\boldsymbol M_\mathrm{Fe2}$ \cite{volkov}.

The suppression of $T_c$ in the S layer of an S/F1/F2 proximity
system studied in our work takes place due to ``leakage'' of Cooper
pairs into the F part. In this language, the generation of the
LRTC at noncollinear magnetizations opens up an additional channel
for this leakage, hence $T_c$ should be suppressed stronger. Note
that the triplet superconducting correlations are generated from
the singlet ones (conversion due to the exchange field), reducing
the amplitude of the singlet component in the S layer and thus
``draining the source'' of superconductivity in the whole system
\cite{comment1}. This effect is substantial since the magnitudes
of the proximity-induced singlet component and the LRTC can be of
the same order near the interface of the S layer (if the thickness
of the adjacent F layer is smaller than its coherence length).

Similarly to the S/F1/F2 system, the LTRC due to noncollinear ferromagnets is also generated in a multilayer of the F1/S/F2
type, studied both theoretically and experimentally \cite{FGK,LCE,Zhu}. At first sight, the same arguments about possibility to additionally suppress
$T_c$ due to leakage of Cooper pairs into the long-range triplet channel are also valid here. However, this suppression is masked by a simple effect
of (partial) mutual compensation of the exchange fields of the two F layers at nonparallel orientations. This effect turns out to be the main one in
the F1/S/F2 geometry, and its monotonicity leads to a monotonic $T_c(\alpha)$ dependence \cite{FGK,LCE,Zhu}. On the other hand, in the S/F1/F2
geometry the outer F2 layer is separated from the S layer, and the mutual compensation of exchange fields has a much weaker effect on the latter. It
is this feature that makes the S/F1/F2 structure advantageous for investigating a nontrivial influence of the triplet component on $T_c(\alpha)$.

The basis of the present work has been formed by our earlier studies of the SC spin valve effect in the multilayer system CoO/Fe1/Cu/Fe2/In \cite{leksin1}. In addition to the observation of the full spin valve effect we have found that its magnitude $\Delta T_c=T_c^\mathrm{AP}-T_c^\mathrm{P}$, i.e., the difference of $T_c$ values for the AP and P mutual orientations of the
magnetizations of the Fe1 and Fe2 layers, shows an oscillating sign-changing behavior as a function of the thickness of the Fe2 layer due to the
interference effect of the SC pairing function in the Fe2 layer \cite{leksin2,leksin3}. As discussed above, the activation of the triplet channel should be visible in an additional suppression of $T_c$ for noncollinear arrangements
of $\boldsymbol M_\mathrm{Fe1}$ and $\boldsymbol M_\mathrm{Fe2}$. Unfortunately, such experiment on the CoO/Fe1/Cu/Fe2/In system turned out to be
unrealizable under well-controlled conditions owing to a low value of $T_c$ for indium and its extreme sensitivity to small out-of-plane tilting of
the external magnetic field. In this respect, lead has much better SC critical parameters, which has determined its choice as an S layer in the
present work.

The CoO/Fe1/Cu/Fe2/Pb samples were prepared by classical electron beam evaporation in an ultra-high vacuum chamber (for details see \cite{leksin3}). In order to obtain a continuous Pb
layer with the resistivity ratio $RRR=R(300\,\mathrm{K})/R(10\,\mathrm{K})>10$ [where $R(300\,\mathrm{K})$ and $R(10\,\mathrm{K})$ are the electrical resistivity
of samples at 300 and 10\,K, respectively], the evaporation rate for the Pb layer was set to 3\,nm/s \cite{comment_evap_rate}.  The easy axis of the magnetization which is induced
by residual magnetic field in our vacuum system was directed parallel to the long side of our rectangular-shaped samples.

We have studied six spin valve samples
CoO$_x$(4\,nm)/Fe(2.5\,nm)/Cu(4\,nm)/Fe($d_\mathrm{Fe2}$)/Pb(35\,nm)
with the thickness of the Fe2 layer $d_\mathrm{Fe2}=0.6$, $0.7$, $0.9$, $1.0$, $1.3$, $1.5$\,nm and one reference
sample CoO$_x$(4\,nm)/Cu(7\,nm)/Fe(1\,nm)/Pb(35\,nm)/
SiN$_x$(85\,nm). The CoO$_x$ layer was used to pin the
magnetization of the adjacent Fe layer (Fe1), whereas the role of
the Cu layer is to decouple the magnetization of the Fe2 layer
from that of the Fe1 layer.

For all samples, we have performed the magnetization measurements
using a VSM SQUID magnetometer. We have measured the major and
minor hysteresis loops $M(H)$ in order to determine the field
range in which the full switching between P and AP orientations of
the magnetizations of the Fe1 and Fe2 layers is realized.
\begin{figure}[t]
\centering{\includegraphics[width=0.8\columnwidth,angle=-90,clip]{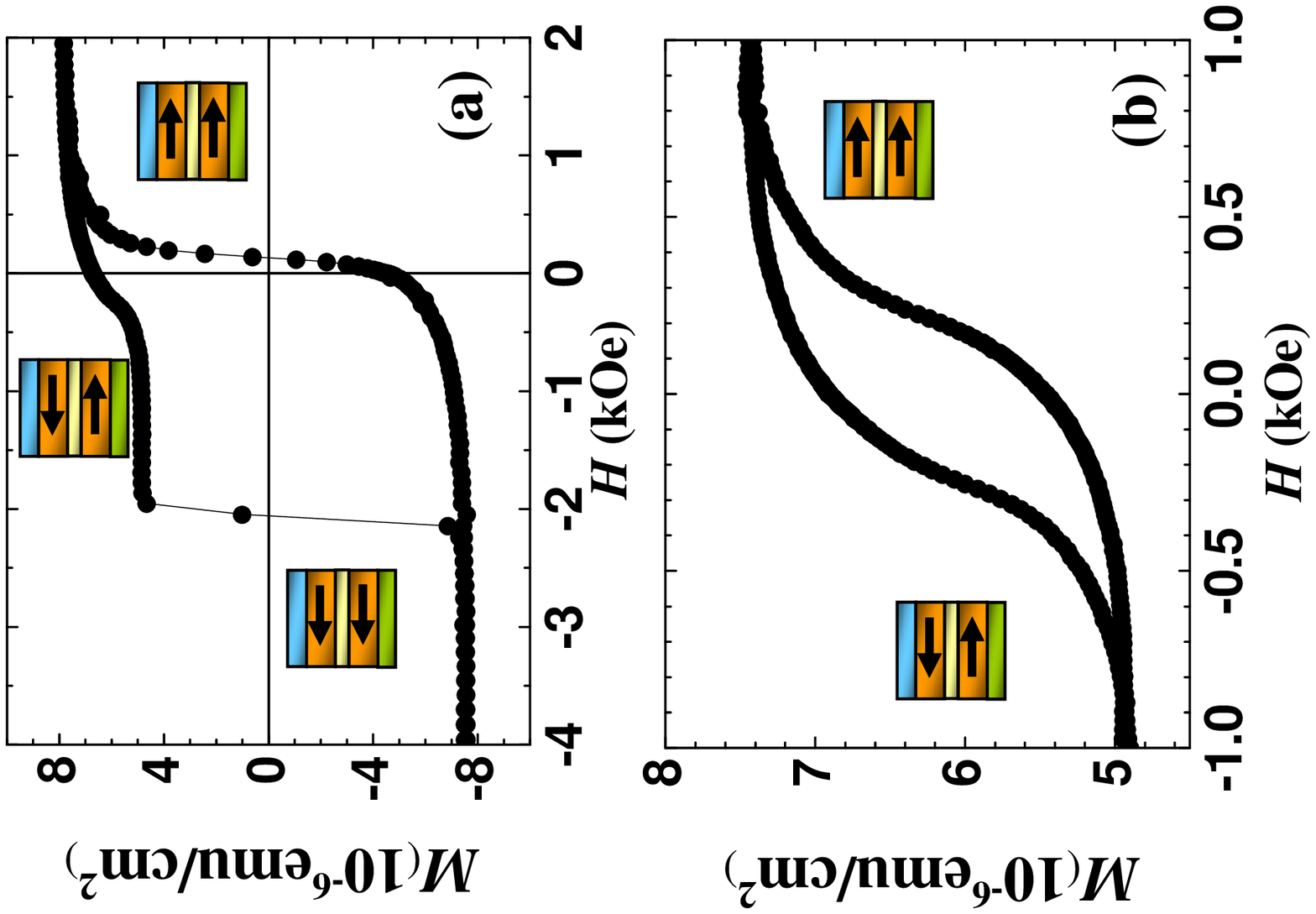}}
\caption{(Color online) (a)~Major magnetic hysteresis loop for the
CoO/Fe1/Cu/Fe2/Pb sample with $d_\mathrm{Fe2}=0.7$\,nm cooled from
room temperature down to $T=4$\,K in a magnetic field $H=4$\,kOe;
(b)~central part of the minor hysteresis loop for the same sample
due to the reversal of the magnetization of the free Fe2 layer.}
\end{figure}
Figure~1 shows a representative magnetic hysteresis loop for the sample with $d_\mathrm{Fe2}=0.7$\,nm which shape is similar for all studied spin valve
samples (see also Ref.~\cite{leksin3} for details). The pinning of the magnetization of the Fe1 layer by the bias CoO$_x$ layer was achieved by
cooling the sample in a magnetic field of $+4$\,kOe applied parallel to the sample plane down to the base temperature $T = 4$\,K at which all
hysteresis loops have been measured. The reversal of $\boldsymbol M_\mathrm{Fe2}$ with respect to $\boldsymbol M_\mathrm{Fe1}$ is illustrated by the minor
hysteresis loop which was obtained by measuring $M(H)$ with decreasing the field from $+4$\,kOe down to $-1$\,kOe and increasing the field again up
to $+1$\,kOe [Fig.~1(b)]. Notably, the minor loop closes and fully saturates at a field of $\pm 1$\,kOe suggesting a complete suppression of the
domain state. We find that the height of the loop $M(\mathrm{P}) - M(\mathrm{AP})$ is proportional to the thickness of the free F2 layer, as expected.

The electrical resistivity measurements were performed with a standard four-point probe setup in the dc mode. The $T_c$ is defined as the midpoint of
the SC transition, and the error bars are related with the transition width. We have combined the electrical setup with a vector magnet that enables a
continuous rotation of the magnetic field in the plane of the sample. To avoid the occurrence of the unwanted out-of-plane component of the external
field the sample plane position was always adjusted with an accuracy better than 3$^{\circ}$ relative to the direction of the dc external field.

Figure~2 shows the dependence of the magnitude of the spin valve
effect $\Delta T_c=T_c^\mathrm{AP}-T_c^\mathrm{P}$ on the
thickness of the Fe2 layer. For samples with $d_\mathrm{Fe2}<0.95$\,nm, we have observed the direct effect with
$T_c^\mathrm{P}<T_c^\mathrm{AP}$, whereas for samples with $d_\mathrm{Fe2}>0.95$\,nm the inverse effect with
$T_c^\mathrm{P}>T_c^\mathrm{AP}$ has been found. Exemplary SC
transition curves for the sample with $d_\mathrm{Fe2}=1.3$\,nm at
$H=1$\,kOe at three different angles $\alpha$ between $\boldsymbol
M_\mathrm{Fe1}$ and $\boldsymbol M_\mathrm{Fe2}$ are shown in the inset
to Fig.~2. The data shown in Fig.~2  can be
explained within the S/F proximity effect in the
framework of the theory of Ref.~\cite{fominov}. In this theory,
the set of equations for the singlet SC component in the S layer contains a real-valued parameter $W(\alpha)$ that enters the boundary
condition for the F2/S interface. Physically, $W(\alpha)$
determines how strongly superconductivity in the S layer is
suppressed by the rest of the structure due to the proximity
effect. The larger $W$ is, the stronger  $T_c$ is
suppressed. With the notion that $T_c$ is a monotonic function
of $W(\alpha)$ \cite{fominov}, we plot in Fig.~2 the theoretical
result for $\Delta W = W(0)-W(\pi)$ together with the experimental
data on $\Delta T_c(d_\mathrm{Fe2})$ (see Ref.~\cite{leksin3} for
details). Figure~2 indeed demonstrates a very good correlation
between $\Delta T_c$ and $\Delta W$ \cite{comment}.
This proves that the
damped oscillating behavior of the spin valve effect is due to the
interference of the Cooper-pair wave functions in the Fe2 layer
\cite{fominov,leksin3}.
\begin{figure}[t]
\centering{\includegraphics[width=0.6\columnwidth,angle=-90,clip]{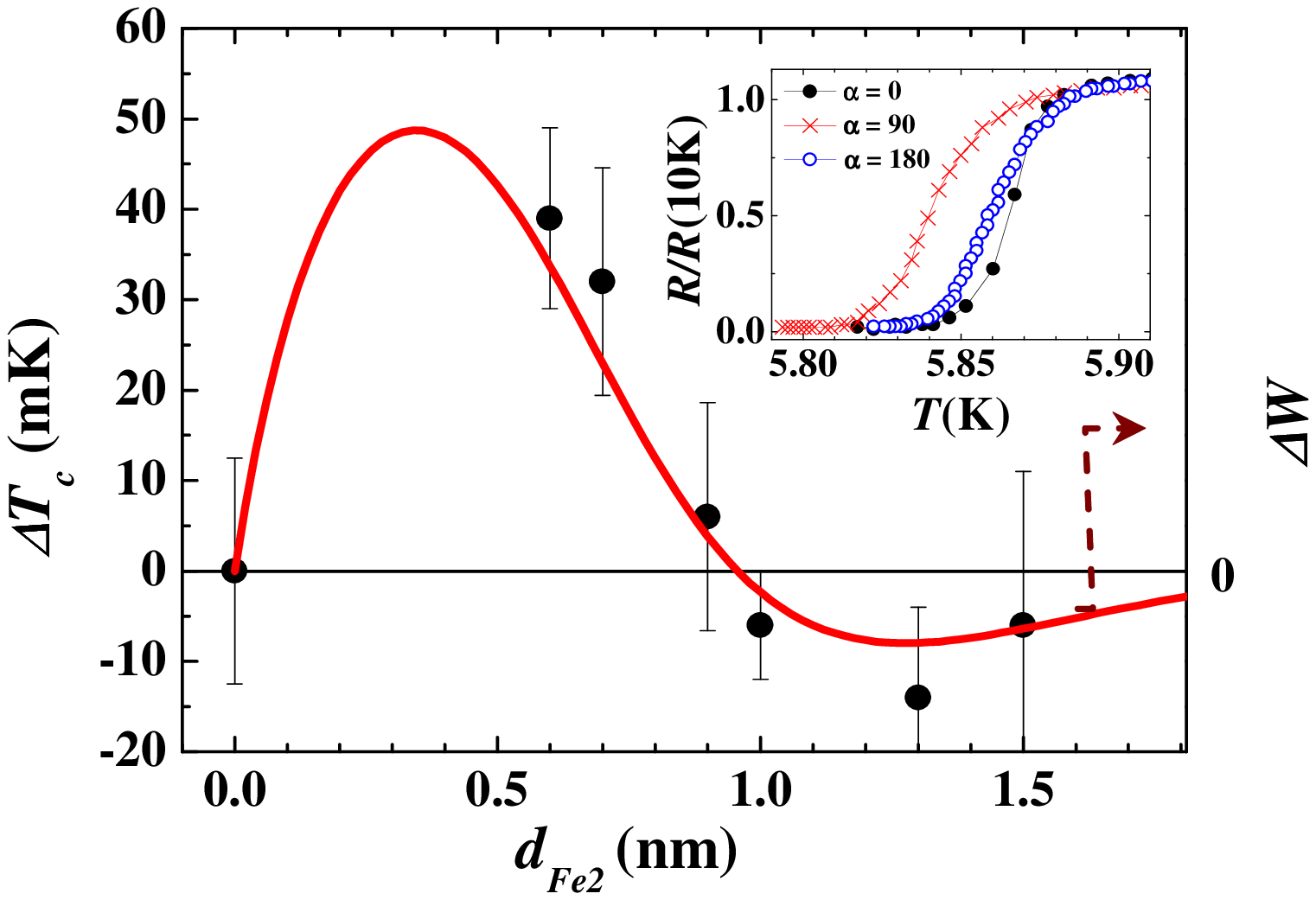}}
\caption{(Color online) Dependence of the magnitude of the spin
valve effect $\Delta T_c$ on the thickness of the Fe2 layer at a
fixed value of the S layer $d_\mathrm{Pb}=35$\,nm. Solid line is a
theoretical curve (see the text). Inset shows the SC transition
curve for the sample with $d_\mathrm{Fe2}=1.3$\,nm at $H=1$\,kOe for
three different angles between magnetizations of the Fe1 and Fe2
layers.}
\end{figure}

The central result of our work is an observation in all spin valve
samples of a very special dependence of the SC critical
temperature $T_c$ on the angle $\alpha$ which the magnetization of
the Fe2 layer $\boldsymbol M_\mathrm{Fe2}$ controlled by an external
field makes with the magnetization of the pinned Fe1 layer.
Examples of such dependences for selected spin-valve samples of
different thickness $d_\mathrm{Fe2}$ are shown in Fig.~3.
\begin{figure}[t]
\centering{\includegraphics[width=0.9\columnwidth,angle=-90,clip]{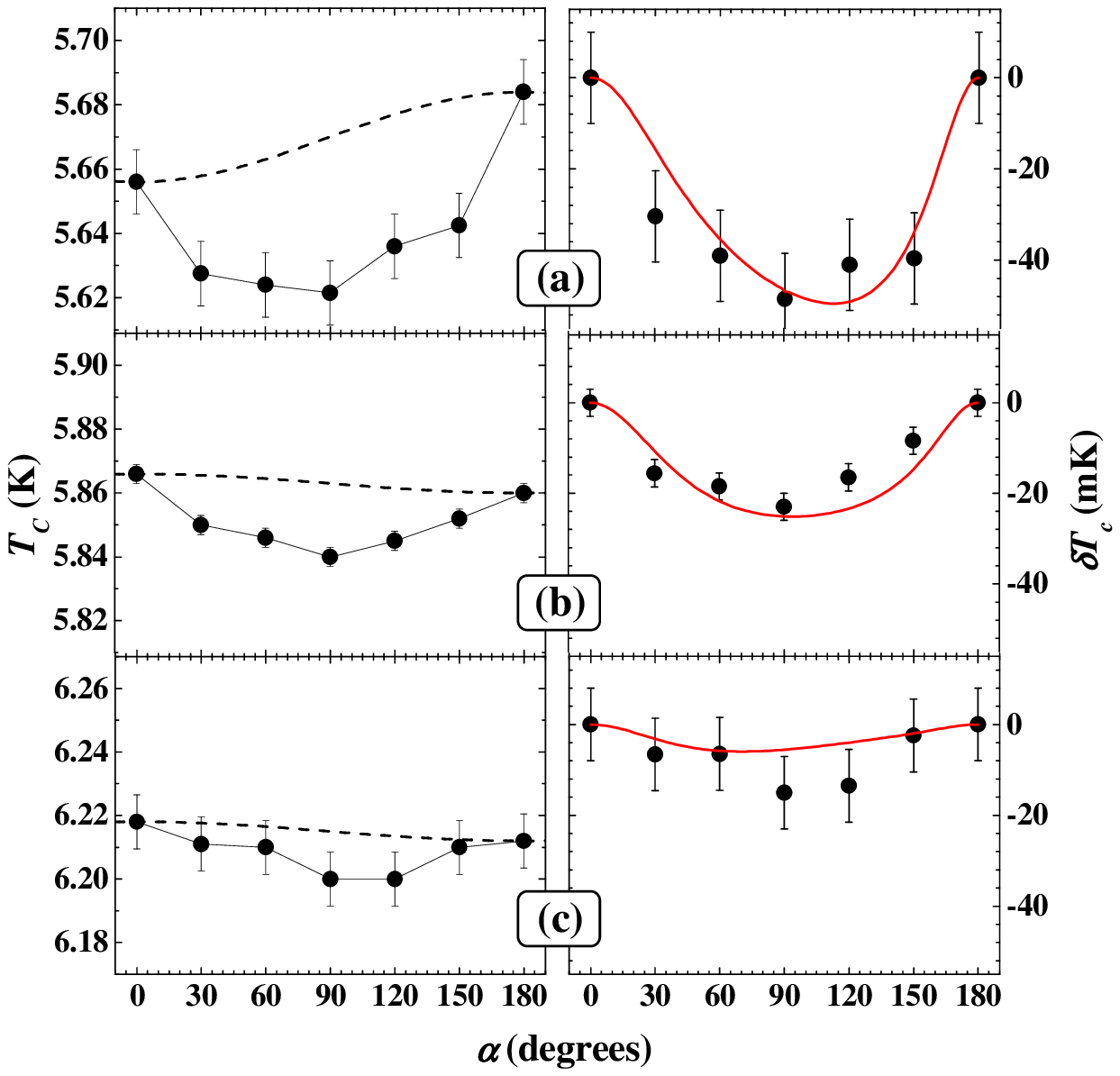}}
\caption{(Color online) Left: dependence of $T_c$ on the angle
between magnetizations of the Fe1 and Fe2 layers measured in a
field $H=1$\,kOe for the samples with $d_\mathrm{Fe2}=0.6$\,nm~(a),
1.0\,nm~(b), and 1.5\,nm~(c). Dashed lines are the reference curves
calculated according to Eq.\ (\ref{Tc_naive}). Right: deviations
$\delta T_c$ of the actual $T_c$ values from the respective
reference curves. Solid lines are theoretical results for $\delta
W$ (see the text).}
\end{figure}
One can see that when changing the mutual orientation of
magnetizations by a gradual in-plane rotation of the magnetic
field from the P ($\alpha =0^\circ$) to the AP ($\alpha
=180^\circ$) state, $T_c$ value does not change monotonically but
passes through a minimum. Importantly, for the reference sample
consisting of just one Fe layer the angular variation of $T_c$
lies within the error bars (not shown). In the following, we argue
that a characteristic minimum in $T_c(\alpha)$ close to $\alpha =
90^\circ$ is a fingerprint of a long-range triplet SC component.
Though the triplet component is inherent in the case of
noncollinear magnetizations, assuming for a moment its absence one
would expect the $T_c(\alpha)$ dependence to be monotonic. From
general symmetry, $T_c(\alpha)$ must behave as $\alpha^2$ and
$(\pi -\alpha)^2$ when $\alpha$ deviates from $0$ and $\pi$,
respectively. One would arrive then at a simple angle-dependent
superposition of the limiting values of $T_c$:
\begin{equation} \label{Tc_naive}
T_c^\mathrm{(ref)}(\alpha)=T_c^\mathrm{P} \cos^2 (\alpha/2) + T_c^\mathrm{AP} \sin^2 (\alpha/2).
\end{equation}
This dependence is shown in Figs.~3(a), (b), and (c) (left side)
by the dashed lines, and we consider them as reference curves.
Deviations $\delta T_c$ of the actual $T_c$ from the reference
curves are, as the figures demonstrate, beyond the experimental
error bars. The angular dependences of this deviation are shown on
the right side panels (a), (b), and (c) of Fig.~3. As discussed
above, in the theory of the triplet spin valve in
Ref.~\cite{fominov} all information about the magnetic part of the
system is encoded into a real positive parameter $W(\alpha)$.
Hence, the $\delta T_c (\alpha)$ dependence should correlate with
deviation of $W(\alpha)$ from a reference curve (increase of $T_c$
corresponds to decrease of $W$); it seems natural to define the
reference curve for $W$ in the same manner as Eq.\ (2), i.e., as
$W^\mathrm{(ref)}(\alpha)=W(0) \cos^2 (\alpha/2) + W(\pi/2) \sin^2
(\alpha/2)$. The explicit expression for $\delta W(\alpha)$ can be
obtained as described in \cite{fominov}, however, the result is
too cumbersome, and we do not present it here. Using the
parameters of the theory \cite{parameters} which we obtained earlier in
Ref.~\cite{leksin3} and taking the appropriate thicknesses of
the Fe2 layer, we find good agreement between the angular dependences of $T_c$ and $W$, see Fig.~3 (right panel).

The dependence of the maximal deviation of $T_c$, which we denote as max\,$\delta
T_c$, on the thickness $d_\mathrm{Fe2}$ is shown in Fig.~4. Since
max\,$\delta T_c$ occurs near  $\alpha =\pi/2$ the dependence
shown in Fig.~4 is close to the dependence of $\delta T_c(\pi/2) =
T_c(\pi/2)-[T_c(0)+T_c(\pi)]/2$ on $d_\mathrm{Fe2}$. As explained above,
this dependence should correlate with $\delta W(\pi/2) =
[W(0)+W(\pi)]/2-W(\pi/2)$ which in accordance to the theory
\cite{fominov} is equal to
\begin{multline}
\frac{\delta W}{W(0)} = -\frac{\sqrt 2 \sin(2k_h d_\mathrm{Fe2} + \pi/4)-
e^{-2k_h d_\mathrm{Fe2}}}{2\left[ \sinh(2k_h d_\mathrm{Fe2}) +\cos(2k_h d_\mathrm{Fe2}) \right]} \\
 - \frac{4 \left[ \sin^2 (k_h d_\mathrm{Fe2}) -2k_\omega d_\mathrm{Fe2}
\right]}{e^{2k_h d_\mathrm{Fe2}}-2+\sqrt 2 \cos(2k_h d_\mathrm{Fe2} +\pi/4) +
4k_\omega/k_h}.
\end{multline}
Here, $k_\omega=\sqrt{2\omega/D}$, $k_h=\sqrt{h/D}$, $\omega=\pi T_c (2n+1)$ with integer $n$ is the Matsubara frequency, $h$ is the exchange field in
the Fe2 layer along the $z$ direction and $D$ is the diffusion coefficient of conduction electrons in the F layer. The result (3) is valid at
$k_\omega \ll k_h$ (obviously satisfied in real ferromagnets with $T_c \ll h$), $k_\omega d_\mathrm{Fe2} \ll 1$ (that is the limit in which the
triplet component is maximal since its spatial damping is negligible), $k_\omega d_S \gtrsim 1$ (this means that the S layer is not too thin and
hence the superconductivity is preserved). The curve corresponding to Eq.\ (3) demonstrates  good agreement with the experimental results (Fig.~4)
confirming that the increase/decrease of $T_c$ indeed correlates with the decrease/increase of $W$. This model curve has been obtained with the same
set of parameters as the ones in Fig.~3 \cite{parameters}.
\begin{figure}[t]
\centering{\includegraphics[width=0.6\columnwidth,angle=-90,clip]{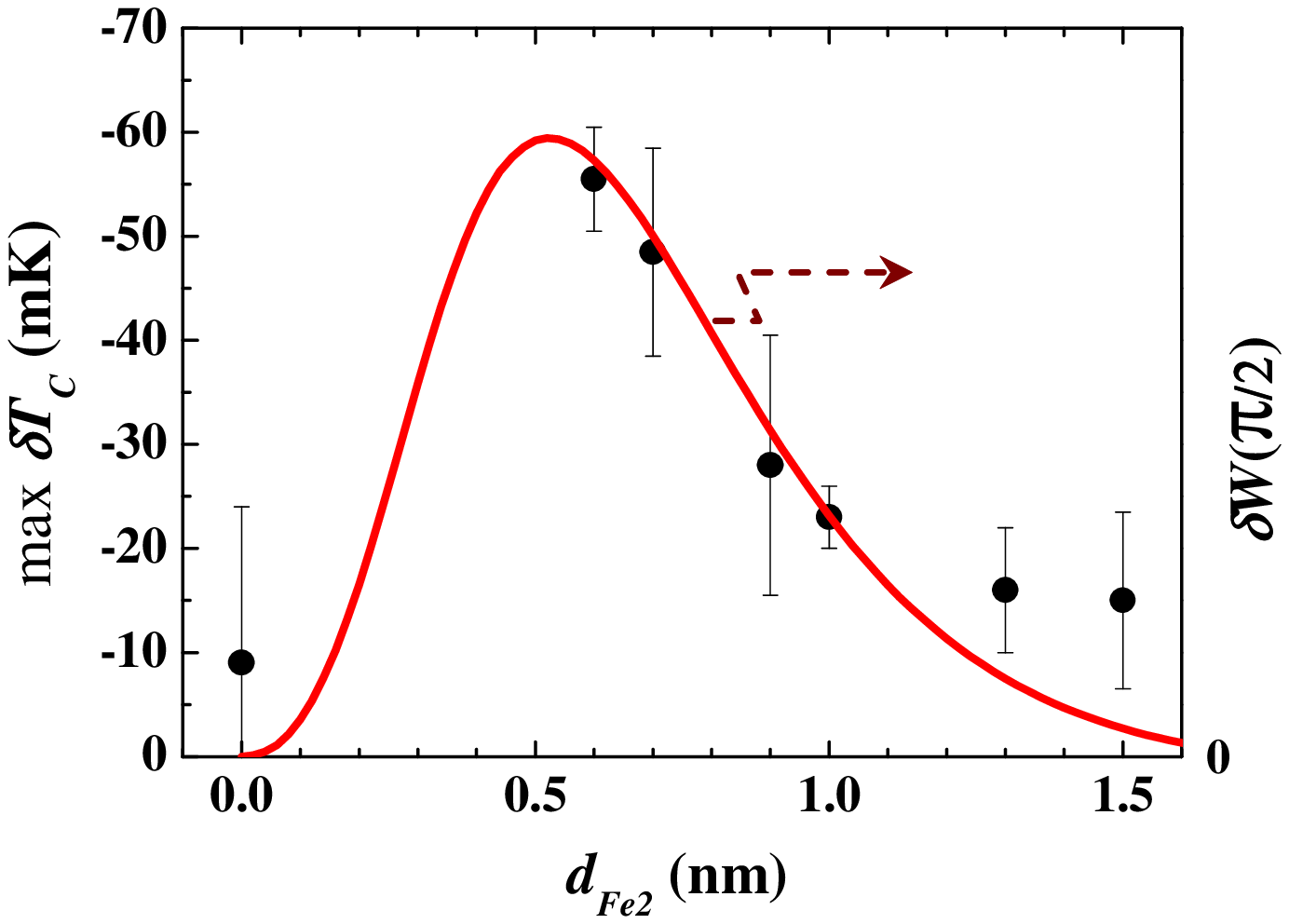}}
\caption{(Color online) Dependence of the maximal deviation of $T_c$,
max\,$\delta T_c$, on the thickness of the Fe2 layer. Solid line is the theoretical result for $\delta W(\pi/2)$ according to Eq.\ (3) (see the
text).}
\end{figure}
Bearing in mind that the effect is not observed for the reference
sample CoO$_x$/Cu/Fe/Pb with a single iron layer we interpret our
finding as evidence for long-range triplet superconductivity that
arises in the spin-valve samples with a noncollinear geometry of
magnetizations of the Fe1 and Fe2 layers.

Finally, we mention that earlier indications for long-range
superconductivity in an F layer have been detected through the
proximity-induced conductance \cite{giroud,petrashov} even before
the theoretical works have appeared. Recently, the occurrence of
the odd in the Matsubara frequency triplet superconductivity in
the S/F/S systems, predicted in Ref. \cite{volkov}, was inferred
from the experiments on Josephson junctions through observation of
the anomalously deep penetration of the Cooper condensate into the
F layer (see, e.g.,
\cite{sosnin,keizer,chan,blamire,aarts,khaire,Westerholt10}).  We note
that our experiments are advantageous in that they address the
primary SC parameter of the spin valve, the behavior of $T_c$,
which is directly affected by the spin-triplet component.

In conclusion, we have studied the behavior of the superconducting
critical temperature $T_c$ of the S/F spin valve system
CoO$_{x}$/Fe1/Cu/Fe2/Pb with different thicknesses of the
ferromagnetic Fe2 layer. We have observed a remarkable
nonmonotonic dependence of $T_c$ on the angle between the
directions of the magnetization $\boldsymbol M$ in the Fe1 and Fe2
layers. The $T_c$ passes through a clear minimum near the
orthogonal orientation of $\boldsymbol M_\mathrm{Fe1}$ and
$\boldsymbol M_\mathrm{Fe2}$ which is not expected in the case of
singlet superconductivity. We argue that this particularly strong
suppression of $T_c$ in the orthogonal geometry is due to an
enhanced ``leakage'' of the SC Cooper pairs into the F layer
occurring via the long-range spin-triplet channel.

\begin{acknowledgments}
The authors are grateful to A.~F.\ Volkov for useful discussions. The work was supported by the RFBR (Grants No.\ 11-02-01063 and No.\ 12-02-91339-NNIO\_a) and by the DFG (Grant No.\
BU 887/13-2). Ya.V.F.\ was supported by the RFBR (Grant No.\ 11-02-00077-a) and the program ``Quantum mesoscopic and disordered structures'' of the RAS.
\end{acknowledgments}


\begin{thebibliography}{99}
\bibitem{comment1}
In the same language, we can explain
stronger suppression of superconductivity in S/F structures,
compared to S/N systems with normal-metallic layers,
as being due to opening of additional triplet channels
(compared to only one singlet channel in the S/N case).
\bibitem{comment_evap_rate}
A high evaporation rate is required to obtain a  homogenous Pb film with a high resistance ratio RRR and a sharp superconducting transition. At lower
rates the immiscibility of Fe and Pb gives rise to an island growth which inhibits the formation of a continuous Pb film on the surface of the Fe
layer (see also \cite{leksin3}).
\bibitem{comment}
Note that although the Usadel equations, employed in
\cite{fominov}, are formally not valid in the case of strong
ferromagnets, it turns out that often they can describe
experimental results surprisingly well (see, e.g., \cite{Birge}).
\end{thebibliography}
\end{document}